\documentstyle [epsf,12pt]{article}
\input epsf
\begin{document}
\newcommand{\be}{\begin{equation}}
\newcommand{\ee}{\end{equation}}
\def\bq{\begin{eqnarray}}
\def\eq{\end{eqnarray}}
\def\ni{\noindent}

\begin{center}
{{\Large Symposium on Cosmology and Astrophysics} \\
\vspace{0.5cm}
29-30 January 2002 \\
\vspace{0.5cm}
Department of Physics \\
Jamia Millia Islamia}
\end{center}

\vspace{1.5cm}

\begin{center}
{{\large Keynote Address}\\
\vspace{0.5cm}
{\sf\large The Relativistic World: A Common Sense Perspective} \\
\vspace{0.5cm}
{\bf Naresh Dadhich}\\
{\it Inter-University Centre for Astronomy and Astrophysics,}\\
{\it Pune, India.}}
\end{center}

\vspace{1cm}
               
\noindent
{\bf The Relativistic World: A common sense perspective} \\

\ni
 First of all I thank the organizers of the workshop for asking me to give 
the keynote address. What is expected of it is perhaps to say something 
profound and interesting which should be accessible to anyone who ventures to 
step into the lecture hall. This is of very tall order for a person like me. 
Since my good friends have posed some confidence in me, I shall try my best 
to live up to their expectations and the occasion. \\

\ni
 I would attempt to present to you a new way of looking at things - a 
farmer's perspective. What I mean by that is to get to the basic root of 
concepts with a minimal degree of mathematical and technical sophistication, 
common sense being the sole guide and torch bearer. That means what I am 
going to talk should in principle be accessible and hence satisfying one of 
the key criteria for a key note address. \\

\ni
 The concepts of space, time and matter are universal and fundamental. A view 
consistent with the contemporary understanding of these basic concepts should 
permeate a rational and truthful world view. By world view I would mean a broad 
enough framework to probe and understand the world around us, the Universe. 
For this a correct and proper understanding of space, time and matter is 
therefore of primary importance. In this lecture, I shall essentially address 
to these concepts and shall attempt to thread a profound synthesis, which is 
given the name, Relativistic World. \\

\ni
 1. The Newtonian World \\

\ni
 In the Newtonian world, space, time and matter are independent and 
externally provided entities. That is, the question of their creation and 
existence is not admitted. Motion is described by the three laws of motion 
governing motion of all that moves. The interaction between bodies in the 
Universe is described by the law of gravitation which prescribes universal 
attraction between all bodies and its measure is proportional to their masses 
and inversely proportional to square of the distance separating them. \\

\ni
The first law of motion makes an equivalence statement that Uniform Motion 
(motion with a constant speed in a straight line) is equivalent to No Motion 
(rest). That is, it is impossible to distinguish between these two states by 
performing any physical experiment. Any deviation from this would signal 
presence or application of an external field or force. In a force free region,
 a particle would either be at rest or in uniform motion depending upon the 
initial condition. It would continue in this state until a force acts on it. 
The only way force makes its presence felt is by changing the velocity of the 
particle it is acting on. \\

\ni
 Conversely if something is moving in a straight line, either there acts no 
force on it or it acts in the same direction of motion. If something moves in 
a straight line in all directions, that means no force is acting on it unless 
there exists in Nature a force which is isotropic - acting in all directions 
at the same time. If there doesn't exist such a force in Nature and yet there 
exists something that moves in a straight line in all directions, its velocity 
must be a universal constant. So long as everyone agrees on its straight line 
motion, its velocity must be the same constant for everyone. \\

\ni
 Now the question arises, does the Newtonian mechanics permit such a universal 
velocity which is the same for all observers irrespective of their relative 
motion? Obviously not because the law of addition of velocity reads as $ w = 
u + v$ where $u$ is the velocity of a particle relative to an observer who is 
moving with velocity $v$ relative to you. According to this law $w$ can never 
be equal to $u$ unless $v = 0$. \\

\ni
 If there exists in Nature a physical entity which propagates in a straight 
line in all directions, it cannot be accommodated in the Newtonian mechanics 
and we will have to seek a new mechanics. \\

\ni
 Inertial frame (IF) is a frame which is free of all forces. It is a measure 
of zero force. All interactions/forces will be measured relative to it. It is 
therefore of fundamental importance to actually realise such a frame. There 
are four basic forces in Nature, two of which are short range nuclear forces 
while electromagnetic and gravitation are long range. One doesn't have to 
worry about nuclear forces, while it is possible to shield electromagnetic 
force by proper prescription of charges and currents but it is impossible to 
shield gravity. Any rearrangement of matter distribution will not be able to 
remove gravity because unlike electromagnetic field which is both attractive 
as well as repulsive, it is always attractive. That means it is impossible to 
realise IF in practice. \\

\ni
 Not quite so. Galileo made a fundamental discovery by dropping two unequal 
mass bodies of different composition that gravity is a truly democratic force 
it pulls everything equally irrespective of its mass and composition. This 
gives a remarkable message that gravity can be removed locally by letting the 
frame fall freely along with everything else. Two freely falling particles will 
have uniform relative motion between them which is the indication of absence 
of force between them. This region could only be local, freely falling lift, 
 because all particles fall towards a point, the centre of force. The extent 
of the force free region is determined by the degree of accuracy required for 
 force freeness. \\

\ni
 We thus come to the conclusion that IF could exist only locally. We can only 
obtain LIFS (local IFs) in actuality and there can exist no global IF. With 
this the first law of motion would state that all LIFs are equivalent for 
investigation of any physical phenomenon. \\

\ni
 Let us once again bring back the thing which has universal constant 
velocity relative to all observers. Gravitation is also universal force and 
hence it must act on everything. The only measure of action of any force is 
the change it produces in velocity of the particle being acted upon. \\

\ni
 The key question is how to make the thing that has universal constant 
velocity feel the gravitational force, because its velocity cannot change. The
incorporation of the universal constant velocity in mechanics asked for new 
mechanics, and now the incorporation of such a particle's interaction with 
gravitation asks for a new theory of gravitation.  \\

\ni
 We have thus argued that the Newtonian world is unsustainable if a particle 
of universal constant velocity does exist in Nature, and we have to seek new 
world which we call Relativistic. \\

\ni
 2. Living with Light \\

\ni
 Light is in fact that thing which is, even in the Newtonian world, supposed 
to move in a straight line in all directions. Hence its velocity must be 
constant for all observers. It is important to note that we have reached this 
important inference of constancy of velocity of light within the Newtonian 
framework without reference to light being an electromagnetic wave propagating
in an all pervading medium called vacuum. And this fact is in direct conflict 
with the Newtonian mechanics. This logical inconsistency had lived with the 
theory over 300 years until the beginning of the previous century. \\

\ni
 2.1 Special Relativity \\

\ni
 The incorporation of universal constant velocity of light gives rise to the 
relativistic mechanics, known as the theory of special relativity. It 
approximates to the Newtonian mechanics when velocities involved are small 
compared to the velocity of light. This was the reason that we could live 
with the inconsistency of principle for so long because we did not encounter 
velocities comparable to velocity of light which could expose the 
inconsistency in observation. Since the velocity of light is a universal 
constant, it binds space and time into a synthetic entity spacetime. That is, 
we cannot confine anything solely to space or to time, and all physical 
happenings happen in spacetime. For instance, distance could be measured in 
seconds, distance traveled by light in so many seconds, and conversely time 
in metres, time taken by light to travel so many metres. \\

\ni
 The immediate consequence of this synthesis is the path dependence of time. 
It is common knowledge that the distance between two points depends upon the 
path you take to go from one point to the other. Since there now exists the 
equivalence between space and time, whatever happens to space measurements 
must happen to time measurements. Consider the time interval between taking 
off a spaceship and its landing back after a space voyage. The time interval 
between these two events would be different in the ground based clock and the 
spaceship borne clock. Since the two clocks take different paths between the 
two events, they must read differently because like distance time interval is 
also path dependent. There is nothing unusual in it except that such a 
situation does not occur in our daily living. That is why it is not so 
familiar. An immediate corollary follows, the absolute simultaneity is 
untenable for spatially separated events because the signal of their 
occurrence would in general reach different observers at different times. The 
events would be simultaneous only for an observer who is equidistant from 
the two events. \\

\ni
 Further it leads to the equivalence of mass and energy indicated by 
the famous equation, $E = Mc^2$, whose practical and devastating power has 
been demonstrated by the atomic bomb. The name of Einstein, the creator of 
the relativistic mechanics, has become synonymous with this equation and to 
his horror and discomfort inseparable from atom bomb. \\

\ni
 At the basic paradigm level we move from 3-dimensional space and 
1-dimensional time to the 4-dimensional spacetime. That is, we live in a 
4- dimensional Universe. This is a major step forward. \\

\ni
 2.2 General Relativity \\

\ni
 Gravitation is universal and hence it must interact with everything 
including light. The question is how to make it interact with light because 
its velocity cannot change. Here we are faced with a contradiction in 
principle. Light must feel gravity yet its velocity must not change. In the 
conventional framework, one can't feel force without its velocity changing.
What is required of light is that it should bend as it grazes close to a 
massive body. Could it be achieved without changing its velocity? \\

\ni
 Consider a piece of wood floating in a river. It bends as the river snakes 
along. Could a similar happen to light. It propagates in space, and the only 
way light can bend is that space bends. That means gravity must bend/curve 
space around a massive body so that light could automatically bend. We have 
now made another very profound discovery that gravitational field must curve 
space, and since space and time have already been synthesized, it must curve 
spacetime. That is the only way to describe gravitation honestly is through 
the curvature of spacetime. A very radical and revolutionary conclusion. The 
study of gravitation through the curvature of spacetime is known as general 
relativity. Once again, this contradiction in principle was not appreciated 
until 1916. There has emerged no significant new fact about gravitation that 
has asked for a new theory. It was only a matter of principle which was true 
even in Newton's time. Had it not been for Einstein's unflinching adherence 
to principle it would have perhaps been not addressed until much later. This 
gives a good measure of Einstein's character and priority in science. \\

\ni
 Note that what the British astronomer, Eddington measured at the time of 
total solar eclipse in 1919 was not the bending of light but rather bending 
of space around the Sun by its own gravity. It is like we see the Sun going 
round the Earth but in fact it is the other way round. Similarly we though 
see the light bending but it is in fact the space bending. Gravity produces 
curvature in spacetime and motion under gravity is simply free motion (of both
 particles and light) relative to the curved spacetime. \\

\ni
  We have now achieved a grander and profounder synthesis of spacetime with 
gravitation which is produced by matter, and hence of spacetime and matter. 
This is something new for physics. No other force makes such a demand on 
spacetime. For all other forces spacetime background is given and it is 
neither affected by nor does it affect. Gravity radically changes this 
paradigm, it works through spacetime curvature. Spacetime no longer remains an 
inert background but becomes dynamic which is essentially the dynamics of 
gravitational field. \\

\ni
 3. Relativistic World \\

\ni
 First the universal character of light velocity led us to binding of space 
and time into spacetime and to the relativistic mechanics called special 
relativity. Next light's interaction with gravity could only be mediated 
through curved space. We can say that light bends space. Gravitation can 
hence only be described by the spacetime curvature and this description is 
called general relativity (GR). It ceases to be an external force but instead 
becomes the property of spacetime geometry. \\

\ni
 Relativistic world is characterized by the synthesis of universal entities 
of space, time, velocity of light and gravitation (matter). The important 
lesson we learn from this is that all that which is universal must be 
synthesized. Never mind even if it requires endowing space and time with 
curvature, a very unusual phenomenon. Spacetime encompasses everything in 
the Universe. That is the nature of spacetime structure is determined by the 
matter energy distribution in the Universe. It is therefore no surprise that 
GR, which is a theory of gravitation as well as of spacetime, 
should admit the question of beginning of time;i.e. beginning or birth of the 
Universe (spacetime). It is a perfectly physics question which of course has 
philosophical overtones. \\

\ni
 The Universe becomes a natural arena for application of GR 
and the study of the Universe of as a whole is called cosmology, which what 
is the subject matter of discussion for this workshop. The two most remarkable 
predictions of the Relativistic World are the beginning of the Universe in an 
explosive Big Bang and the death of a massive star in a black hole. A black 
hole is a highly compact object producing very high field which in turn curves
 space around it so confiningly that even light cannot propagate out of it 
(Recall that light can only move along the space curvature). Since light 
cannot come out of black hole, nothing else could as well. It is an object 
whose boundary surface has turned one way, things can go into it but nothing 
can come out. No information can come out from inside of it. \\

\ni
 When a massive enough star has exhausted its nuclear fuel so that its own 
gravity could not be checked, it shrinks without limit to form a black hole. 
It would contain inside it a singular state where physical parameters become 
infinite. Similar is the case at the big bang. \\

\ni
 4. End of everything \\

\ni
 The breakdown of theory is called by the beautiful name of singularity. It 
indicates a situation where physical parameters of the theory become untenable
 by attaining infinite values. For singularity in GR would 
mean spacetime curvature along with energy density becoming infinite. It would 
mark breakdown of spacetime curvature. The singularity in GR not only marks 
the end of GR, but also of everything else. No other physical theory or 
structure can survive in absence of the proper spacetime background. Thus 
like gravity its singularity is also universal. \\

\ni
 5. Looking ahead \\

\ni
 Gravity is a universal interaction and it is therefore not a force like 
other forces but is a property of spacetime structure of the Universe. The 
question is how do we address the singularity in the structure of spacetime, 
indicating its breakdown not only for description of gravity but also as 
background for all physical phenomena. It indicates the end of everything, 
and should the answer of it be the Theory of Everything? So do the proponents 
of the string theory claim. The string theory is one of the proposals for a 
covering theory to 
GR which is vigorously being pursued. It proposes in principle 
a grand vision of synthesis of all forces, quantization of gravity and origin 
of matter in the Universe. All this is supposed to happen in 10 dimensions. 
The extra space dimensions are supposed to be compact and manifest only at a 
very high energies. GR would be the low energy limit of the theory. There is 
however no natural and unique way to come from 10 to 4 dimensions. \\

\ni
 The other approach is that of canonical quantum gravity a la Abhay Ashtekar 
et al. How to tailor spacetime geometry so that it becomes 
amenable to quantization? It is essentially the quantization of spacetime 
geometry. New mathematical tools and constructs as well as new concepts are 
being developed so as to handle discretization of inherently continuum 
structure of spacetime. They are in fact evolving quantum geometry. This 
proposal remains confined to the familiar 4-dimensions.  \\

\ni
 Both the approaches have about equal share of success, for instance in 
understanding black hole entropy to some extent. Yet they have both long 
way to go. The string theory has a strong backup of the large particle physics 
community while the canonical quantization is pursued mainly by a small 
group of relativists. The two approaches are quite different, the former 
relies on the field theoretic techniques and concepts while the latter on the 
geometric constructs. The two will have to converge as they asymptotically 
approach the ultimate theory of spacetime, which may or may not be a theory 
of everything! At any rate, one thing is certain that we do need a new theory 
which could be string theory or canonical quantum gravity, or something else! \\

\ni
 Let us try to see into the unknown on the basis of wisdom gained from our 
past experience. From Newton to Einstein, the guiding force was light, its 
constant speed and its interaction with gravity. The latter required space 
to be curved. On the basis of these simple facts, we could argue that 
gravitational potential should not only give the acceleration to ordinary 
particles but also curve space for photons to feel gravity. Since acceleration 
is given by gradient, constant potential has no dynamical effect. Photons 
through space curvature directly feel the potential and hence they are 
sensitive to its value. Therefore constant potential is not ignorable as it 
does produce non-zero curvature. The Einstein theory has thus to determine 
in contrast to the Newtonian theory potential absolutely. The journey from 
old to new theory is always of synthesis. It is the good old Newtonian 
potential with its zero fixed at infinity. This is a new synthesis. \\

\ni
 Further, as there exists the escape velocity threshold for timelike 
particles, there should also exist a similar threshold for photons. Photons 
cannot turn back like ordinary particles, the only way they could be kept 
bound to a gravitating body is that they are not let to propagate out of a 
compact surface. Since photons define the limiting threshold for propagation 
of any physical information, from such a surface nothing can come out. It 
then becomes a oneway surface defining the horizon of a black hole. The 
existence of black hole then becomes a natural requirement the moment we have 
photons to contain. \\

\ni
 Such profound and distinguishing features of GR could be deduced in 
principle without reference to the full theory. \\

\ni
 The point I wish to make is that it should always be possible to anticipate 
and deduce some of the features of the new theory. Now when we wish to go 
beyond Einstein in the high energy regime, what kind of new features could one 
expect? Unfortunately, there does not seem to be anything like light showing 
us the path. One possible entity for synthesis is gravitational field energy 
like the gravitational potential in the Newton-Einstein synergy. In GR, it 
only manifests through the space curvature. In the high energy limit, could 
it happen that it becomes concrete like other matter fields. Perhaps it may be 
the right direction to look for. \\

\ni
  The measure of a theory is the kind of questions it admits. New theory admits
questions that were not admissible in the earlier theory. For instance, it is 
a valid question to ask in GR, when did the spacetime (the Universe) begin? 
And the answer is that it had its explosive birth in hot big-bang about 
15-20 billion years ago. The next question in this series for the new theory 
to answer would be, what is the spacetime made of? That is the fundamental 
question which should be answered by the new theory. \\
 
\ni
 A fundamental theory also has an impact on the world view. The view that has 
gained acceptance amongst people at large. For instance, the fact that the 
earth is not flat but is a curved surface like a ball has been internalized 
and assimilated in the knowledge base of the present day society. At a slightly
 higher degree of abstraction, the fact that material bodies attract 
each-other by an invisible force of gravity has also descended down to the 
common knowledge culture. The next step of advancement in the similar vein 
would be the assimilation of the fact that massive bodies curve the space 
around them. That is the abstract spacetime manifold (the Universe) we live 
in is not flat but is curved. One of the ways to measure this curvature is by  
measuring  bending of light when it grazes a massive body like a star. Thus 
the geometry of spacetime should be of as general a concern as the geometry 
of the planet earth we live on. The only difference is that the former is an 
abstract entity while the latter is the concrete rock and sand. \\

\ni
 Like gravity social interaction is also universal and of great consequence 
for our existence and well being. It should thus also attract collective 
concern and attention of all of us with utmost seriousness and commitment. We 
should never forget the fellow citizens who have been contributing from their
honest earnings for our upkeep as well as for the facilities we use for our 
work. Apart from contributing to the knowledge base by our work in our 
specific discipline, as persons of learning and more importantly 
practitioners of scientific method we also owe to the society a studied and 
responsible participation in the discourses on the issues of wider social 
relevance. \\

\ni
 Let us try to emulate the basic gravitational property of interaction with 
one and all, and come closer. If this is realised in the true spirit of both 
science and society, there is a good reason for hope for harmony, peace and 
an enlightened world community. This is a matter of gravity for one and all, 
once again a universal pronouncement of profound significance and value. \\

\ni
 I close by quoting Ghalib, the great Urdu poet of the 18th century. \\

\vspace{0.2cm}

\begin{figure}[h]
\begin{center}
\input{epsf}

\epsfysize = 2.5in

\epsffile{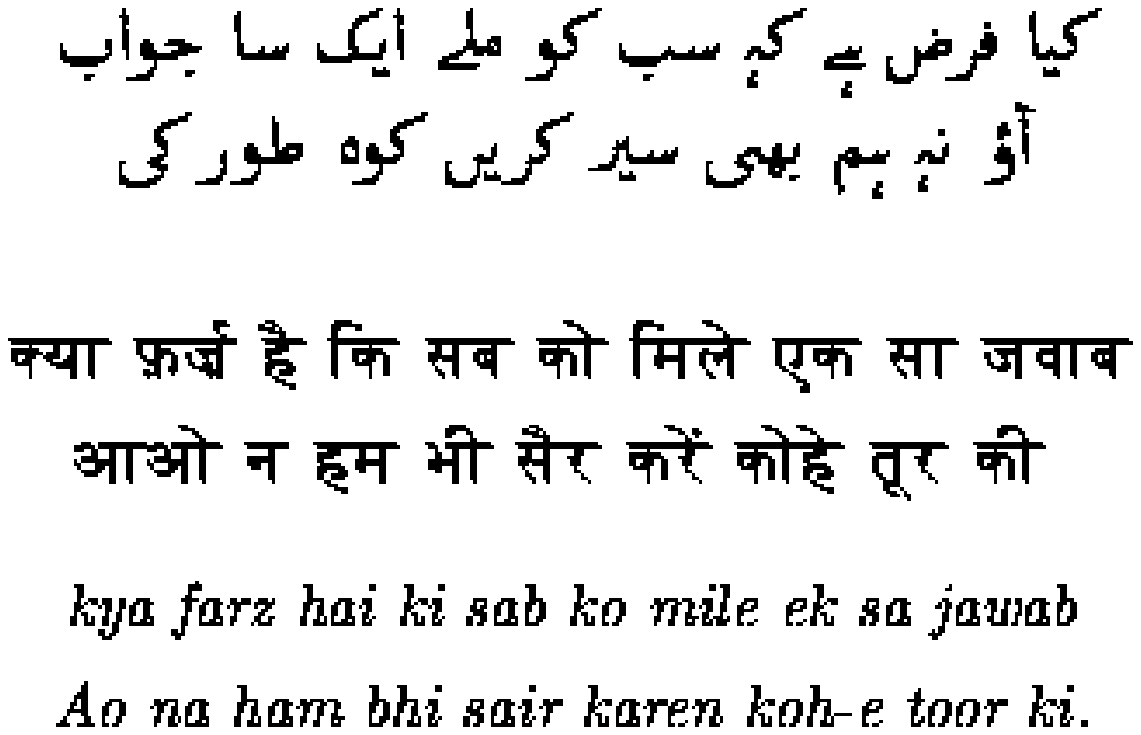} 
\end{center}
\end{figure}

\end{document}